\documentclass[sigconf]{acmart}

\AtBeginDocument{%
 }

\copyrightyear{2025} 
\acmYear{2025} 
\setcopyright{cc} 
\setcctype{by} 
\acmConference[SIGIR-AP 2025]{Proceedings of the 2025 Annual International ACM SIGIR Conference on Research and Development in Information Retrieval in the Asia Pacific Region}{December 7--10, 2025}{Xi'an, China}
\acmBooktitle{Proceedings of the 2025 Annual International ACM SIGIR Conference on Research and Development in Information Retrieval in the Asia Pacific Region (SIGIR-AP 2025), December 7--10, 2025, Xi'an, China}
\acmDOI{10.1145/3767695.3769507} 
\acmISBN{979-8-4007-2218-9/2025/12}

\usepackage{graphicx}
\usepackage{subcaption}
\usepackage{balance}
\usepackage[english]{babel}

\usepackage{microtype}

\begin{document}

\title{Reproducing and Extending Causal Insights Into Term Frequency Computation in Neural Rankers}

\author{Cile van Marken}
\affiliation{%
 \institution{University of Amsterdam}
 \city{Amsterdam}
 \country{The Netherlands}}
\email{cile.van.marken@student.uva.nl}

\author{Roxana Petcu}
\affiliation{%
 \institution{University of Amsterdam}
 \city{Amsterdam}
 \country{The Netherlands}}
\email{r.m.petcu@uva.nl}

\renewcommand{\shortauthors}{Cile van Marken and Roxana Petcu}

\begin{abstract}
Neural ranking models have shown outstanding performance across a variety of tasks, such as document retrieval, re-ranking, question answering and conversational retrieval. However, the inner decision process of these models remains largely unclear, especially as models increase in size. Most interpretability approaches, such as probing, focus on correlational insights rather than establishing causal relationships. The paper \textit{'Axiomatic Causal Interventions for Reverse Engineering Relevance Computation in Neural Retrieval Models'} by \citet{chen2024axiomatic} addresses this gap by introducing a framework for activation patching - a causal interpretability method - in the information retrieval domain, offering insights into how neural retrieval models compute document relevance. The study demonstrates that neural ranking models not only capture term-frequency information, but also that these representations can be localized to specific components of the model, such as individual attention heads or layers. This paper aims to reproduce the findings by \citet{chen2024axiomatic} and to further explore the presence of pre-defined retrieval axioms in neural IR models. We validate the main claims made by \citet{chen2024axiomatic}, and extend the framework to include an additional term-frequency axiom, which states that the impact of increasing query term frequency on document ranking diminishes as the frequency becomes higher. We successfully identify a group of attention heads that encode this axiom and analyze their behavior to give insight into the inner decision-making process of neural ranking models.\footnote{The code and data are available on \href{https://github.com/cilevanmarken/axiomatic_analysis_in_neural_rankers.git}{github}.} 
\end{abstract}

\ccsdesc[500]{Information systems~Retrieval models and ranking}

\keywords{Interpretability, Neural Ranking Models, Information Retrieval Axioms, Search, Activation Patching, Perturbations, Term Frequency, Injections, Attention, Transformers}

\received{1 July 2025}

\maketitle

\section{Introduction}
Information retrieval (IR) systems have become increasingly dependent on neural architectures - particularly transformer-based models - due to their strong performance in capturing complex semantic relationships between queries and documents \cite{karpukhin2020dense, khattab2020colbert}. However, unlike traditional systems that rely on lexical matching, neural models encode information in highly dimensional representation spaces, making them inherently non-interpretable \cite{mitra2018intro}. This is especially challenging as the model increases in size \cite{fernando2019study}. Hybrid approaches such as SPLADE \cite{formal2021splade} offer an alternative by introducing interpretability through sparse representations while learning rich representational spaces. However, fully understanding the decision process of retrieval models remains an open problem, which could enhance reliability and aid in mitigating biases present in the training data.

Axiomatic IR offers a formal framework for analyzing and shaping model behavior based on well-defined properties, which described desirable decision-making guidelines for IR models. One common approach for analyzing whether such properties are encoded in the model is probing, which has provided valuable insights into the behavior of the transformer architecture and how it generates language \cite{ferrando2024primer, poinsot2024learning}. However, probing reveals correlations rather than establish causal relations \cite{chowdhury2024probing}. 

To overcome these limitations causal intervention-based methods, based on causal mediation analysis \cite{pearl2022direct}, have emerged as more robust approaches to understanding how models encode features \cite{zhang2021causal, vig2020gender}. An example of a causal intervention method is activation patching \cite{zhang2023towards}, which modifies specific activations in a model by substituting them with activations from a controlled source, allowing for a systematic analysis of the model's components and their role in the decision-making process of the model. This form of analysis provides a more granular understanding of individual model component contributions compared to probing.

The paper \textit{'Axiomatic Causal Interventions for Reverse Engineering Relevance Computation in Neural Retrieval Models'} by \citet{chen2024axiomatic} combines traditional IR axioms with modern causal intervention-based methods to identify whether neural ranking models encode core information for computing document relevance. Specifically, the authors study the encoding of a core IR axiom, namely TFC1 \cite{fang2004heur}, which describes the expected behavior of IR models with respect to term frequency overlap between queries and documents. The paper introduces activation patching in a novel retrieval setup and establishes practices for creating diagnostic IR datasets for activation patching experiments. \citet{chen2024axiomatic} show that the TAS-B model, as defined by \citet{hofstatter2021tasb}, encodes term-frequency information consistent with the TFC1 axiom, and are able to localize this feature to specific attention heads in the model.

This study investigates the reproducibility of \textit{'Axiomatic Causal Interventions for Reverse Engineering Relevance Computation in Neural Retrieval Models'} and extends it by exploring an additional retrieval axiom TFC2 \cite{fang2004heur}, which states that the impact of query term frequency on ranking decreases with higher frequencies. We are able to validate that TAS-B \cite{hofstatter2021tasb} tracks frequency information and that this behavior can be localized to specific attention heads. By refining the baseline setup, we are able to achieve an even more precise localization of term-frequency information.

For the extension analysis, we propose diagnostic datasets based on practices of \citet{chen2024axiomatic} and conduct experiments for the TFC2 axiom \cite{fang2004heur}. This study makes the following contributions:
\begin{enumerate}
 \item A reproducibility study involving revisions of the original code base, which results in a more precise localization of term-frequency information;
 \item Introducing an experimental setup for retrieval axiom TFC2 within the existing activation patching framework;
 \item Discovering that TAS-B encodes a latent mechanism for tracking term frequencies consistent with the TFC2 axiom.
\end{enumerate}

\section{Related Work}

\subsection{Axiomatic IR}
Traditional IR models are based on explicit rules and axioms, which define specific properties that an effective ranking model should satisfy. Models such as TF-IDF and BM25 adhere to these principles by leveraging properties such as term frequency, inverse document frequency, and document length normalization \cite{robertson2009probabilistic}. Since their first mention in 1994 \cite{bruza1994axiom}, IR axioms have become fundamental tools in information retrieval research as guiding principles that help align model behavior with human expectations for relevance assessment. Neural ranking models are difficult to interpret. However, recent efforts in explainable IR (XIR) have used these axioms to analyze the inner decision processes of neural ranking models. For example, \citet{camara2020heuristics} investigated the extent to which decisions made by neural ranking models can be explained by retrieval axioms, and \citet{rosset2019regularizing} demonstrate that regularizing the training of such models using axioms leads to faster convergence and better generalization. 

\subsection{Probing}
Probing is a widely adopted method for assessing whether neural models have grasped specific concepts as well as identifying in which components of the model these concepts are encoded. It involves training a lightweight classifier on model components, such as embeddings or attention maps, to evaluate whether the concept of interest is present \cite{formal2021white, vaswani2017attention, formal2022match}. Although these methods reveal correlations between model representations and specific concepts, their reliability is debated, as probing does not establish causal relationships \cite{belinkov2022probing, belinkov2019survey}. \citet{chen2024axiomatic} therefore opt for a different approach, involving causal intervention-based methods.

\subsection{Activation Patching}
A widely used causal intervention-based method is activation patching \cite{vig2020gender, meng2022locating, geiger2021causal}, which replaces the activation values of a model component from one forward pass with those from a controlled source. By iteratively patching activations and evaluating the effect on the model logits, one can localize which component is responsible for the behavior related to the task. In practice, activation patching has proven to be effective in detecting gender bias \cite{vig2020gender}, localizing where models store factual information \cite{geva2023factual, macavaney2022behavior} and correcting model errors through editing \cite{sakarvadia2024memoryinjectionscorrectingmultihop}.

\section{Methodology}

\subsection{Axiomatic Analysis in Neural Rankers}

\subsubsection{Activation Patching for IR}
Activation patching methods analyze model behavior by comparing a ‘clean’ input (\textit{X\_clean}), and a ‘corrupted’ input (\textit{X\_corrupted}). The two inputs should be similar, but different enough such that the feature of interest can be isolated in the model and identified in the output. For example, to investigate one-hop reasoning \cite{chen2019hop}, \textit{X\_clean} could be 'Paris is the capital of with answer: 'France', while \textit{X\_corrupted} could be 'London is the capital of' with answer: 'England'. When doing a forward pass with \textit{X\_clean}, the intermediate activations, such as MLP outputs and hidden states, are stored. These can be used to replace the existing activations of the forward pass on \textit{X\_corrupted} in the same component. The impact of the change is evaluated using the model logits. To make activation patching suitable for IR experiments, \citet{chen2024axiomatic} propose the following modifications to the general activation patching setup:

\begin{enumerate}
 \item Instead of using a \textit{X\_clean} and \textit{X\_corrupted} pair, a diagnostic dataset is created using a query, a ‘clean’ document (\textit{X\_baseline}) and a ‘corrupted’ document (\textit{X\_perturbed}). More details about these diagnostic datasets can be found in Section \ref{section:datasets}.
 \item Instead of evaluating the logits of the model, an assessment of the patch’s effect will be done by evaluating the normalized difference in ranking scores \cite{wang2022interpretability} between the baseline and corrupted pairs. As stated by \citet{chen2024axiomatic}, a value of 1 indicates that the intervention increases the ranking score so that it fully recovers the performance of the document with the highest ranking. A value of 0 indicates that the patch had no effect on performance.
\end{enumerate}

The activation patching procedure involves running the model on different input variations to examine the effect of specific activations on ranking behavior:
\begin{enumerate}
 \item \textit{Baseline run}: Forward pass using \textit{X\_baseline} and record the ranking score.
 \item \textit{Perturbed run}: Forward pass using \textit{X\_perturbed}. Store cache activations and record the ranking score.
 \item \textit{Patched run}: Forward pass using \textit{X\_baseline}, replacing a specific activation with the cached values from the perturbed run, and record the ranking score.
\end{enumerate}

\subsubsection{Axiom and Perturbations}
\label{section:TFC1}
Using activation patching, \citet{chen2024axiomatic} analyze whether neural ranking models encode the TFC1 axiom. TFC1 is a term-frequency axiom that connects document relevance to the overlap of terms between query and document and is defined by \citet{fang2004heur}:

\vspace{1mm}
\noindent \makebox[5em][l]{TFC1}%
\parbox[t]{\dimexpr\linewidth-5.5em}{%
 Let $q = w$ be a query with only one term $w$. Assume the length of document $d_1$ equals the length of document $d_2$. If the number of occurrences of $w$ in $d_1$ is greater than the number of occurrences of $w$ in $d_2$, then for query q the relevance score of $d_1$ should be higher than $d_2$.
}
\vspace{1mm}

\citet{chen2024axiomatic} introduce two variants on the TFC1 axiom, namely TFC1-Inject (TFC1-I) and TFC1-Replace (TFC1-R), which are defined as follows:

\vspace{1mm}
\noindent \makebox[5em][l]{TFC1-I}%
\parbox[t]{\dimexpr\linewidth-5.5em}{%
 A term is sampled from the query and inserted at the end of the document $d$ to create perturbed document $d_p$. To create a baseline document $d_b$ equal in length to perturbed document $d_p$, filler token(s) (e.g., ‘a’) are inserted at the end of document $d$.
}
\vspace{1mm}

\noindent \makebox[5em][l]{TFC1-R}%
\parbox[t]{\dimexpr\linewidth-5.5em}{%
 A term is sampled from the query. All occurrences of the sampled query term in document $d$ are replaced with a filler token to create perturbed document $d_p$. The original document $d$ acts as the baseline document $d_b$.
}
\vspace{1mm}

To measure the impact of the location of term-injection, \citet{chen2024axiomatic} separate the TFC1-I experiment into TFC1-I append, where the sampled query term is concatenated to the original document in the last position, and TFC1-I prepend, where the sampled query term is concatenated to the original document at the beginning of the document. The term that is either injected or replaced is the term from the query that causes the highest average change in ranking score after perturbing the documents. This prevents low-IDF terms from being sampled, which could lead to suboptimal diagnostic datasets.

\subsubsection{Diagnostic Datasets}
\label{section:datasets}
For the experiments, diagnostic datasets were created using Passage-Retrieval MS-MARCO \cite{nguyen2016ms}, which contains approximately 6.8k queries. For each query, the top 100 most relevant documents were retrieved and perturbed.

The diagnostic dataset for TFC1-I contains:

\begin{enumerate}
 \item \textit{Baseline document:} The original document with a filler token appended to match the length of the perturbed document.
 \item \textit{Perturbed document:} The original document with the selected query term appended.
\end{enumerate}

For TFC1-R, the dataset contains:
\begin{enumerate}
 \item \textit{Baseline document:} No changes are made to the original document.
 \item \textit{Perturbed document:} All occurrences of the selected query term are replaced with a filler token.
\end{enumerate}

The 100 queries with the highest change in the average retrieval score between the baseline and perturbed documents were selected for experiments, as these queries would best illustrate the impact of activation patching in specific components of the model. To analyze which tokens have the most impact on model performance across different documents, the document's tokens are organized into categories. An overview of the token categories is shown in Table \ref{tab:categories}. 

\begin{table}
 \caption{Token categories\protect}
 \label{tab:categories}
 \begin{tabular}{p{0.15\linewidth} p{0.75\linewidth}}
 \toprule
 Label & Definition \\
 \midrule
 $tok_{CLS}$ & The CLS token. \\
 $tok_{inj}$ & The selected query term injected into the document. \\
 $tok_{qterm+}$ & Occurrences of the selected query term that already
exist in the original document.\\
 $tok_{qterm-}$ & Occurrences of the non-selected query terms in the
original document. \\
 $tok_{other}$ & Terms in the original document that are not query
terms. \\
 $tok_{SEP}$ & The SEP token.\\
 \bottomrule
\end{tabular}
\end{table}

\subsubsection{Experimental Setup}
\label{section:experimentalTFC1}
All experiments are conducted using the TAS-B model, which has 6 layers and 12 attention heads per layer \cite{hofstatter2021tasb}. This model is a highly effective neural ranking model, with a simple architecture that requires less computational resources compared to models such as ANCE and RocketQA \cite{xiong2020approximate, yingqi2020rocketQA}. Activation patching requires iterative interventions on model components with multiple runs per input and is therefore computationally intensive. In addition, the smaller architecture allows for localizing the specific attention heads that are most impacted by the interventions more efficiently.

To ensure consistent document length between the baseline and perturbed documents, the filler token 'a' is used, as it has neutral word semantics and does not change the grammatical structure of the document. The formula for computing the relevance score is illustrated in Equation \ref{eq:relevance_score} and has been used as logit difference by \citet{wang2022interpretability}. 

\begin{equation}\label{eq:relevance_score}
 \frac{\text{patched score} - \text{baseline score}}{\text{perturbed score} - \text{baseline score}}
\end{equation}

\subsection{Axiomatic Extension: Incremental TF Effects on Relevance}
\citet{fang2004heur} propose multiple axioms, including two that relate to term frequency. The TFC1 axiom establishes a fundamental relationship between term frequency and document relevance, whereas the TFC2 axiom introduces an extra constraint: the term frequency impact should decrease as the number of term occurrences grows. Studying the TFC2 axiom can provide a deeper understanding about the model's reasoning choices, while also offering insights into whether TFC1 generalizes when the query term already exists in the document at multiple positions.

\subsubsection{Axiom and Perturbations}
\label{section:TFC2axiom}
The TFC2 axiom \cite{fang2004heur} is defined as follows:

\vspace{1mm}
\noindent \makebox[5em][l]{TFC2}%
\parbox[t]{\dimexpr\linewidth-5.5em}{%
 Let $q = w$ be a query with one term $w$. Assume the length of documents $d_1$, $d_2$ and $d_3$ is equal and $d_1$ contains w at least once. If $TF(w, d_2) - TF(w, d_1) = 1$ and $TF(w, d_3) - TF(w, d_2) = 1$, then the difference in relevance scores $f(q, d_2) - f(q, d_1)$ should be larger than $f(q, d_3) - f(q, d_2)$.
}
\vspace{1mm}

In other words, the TFC2 axiom defines that relevance score increases with term frequency, but that this relation is sublinear: larger term frequencies will result in a smaller increase in relevance scores. Additionally, when two documents have the same number of query term occurrences, the axiom states that the document containing more distinct query terms should be favored.

\subsubsection{Diagnostic Datasets}
Passage-Retrieval MS-MARCO \cite{hofstatter2021teaching} is used to create diagnostic datasets for the TFC2 experiments. We iteratively add a query term and measure the corresponding increase in relevance. The number of insertions is denoted as $K$, with experiments carried out for values of $K$ ranging from 1 to 10. We refer to these configurations as TFC2-K1 through TFC2-K10. For each query, the top 100 retrieved documents are perturbed as follows:

\begin{enumerate}
 \item \textit{Baseline document:} The selected query term is appended to the document $K$ times, as well as a filler term to match the length of the perturbed documents.
 \item \textit{Perturbed document:} The selected query term is appended to the document $K + 1$ times.
\end{enumerate}

\subsubsection{Experimental Setup}
\label{section:TFC2experimental}
We use the TAS-B model \cite{hofstatter2021tasb}, consistent with the original study. As mentioned in Section \ref{section:datasets}, \citet{chen2024axiomatic} select the 100 MS-MARCO queries with the highest change in the average retrieval score between the baseline and the perturbed documents for a given query. The same queries from the TFC1 experiments are used in the TFC2 experiments. Similarly, the query terms for creating diagnostic datasets for the TFC2 experiments \ref{section:datasets} are adopted from the original study. The potential methodological implications of this selection are discussed in Section \ref{section:implicationsdataset}.

For the TFC2 experiments, $K=10$ was chosen as the upper bound for adding query terms to the document. Preliminary experiments up to $K = 50$ indicated that TFC2 behavior emerges for lower values of $K$ and that the increase in relevance scores becomes very small (<0.5\%) for values higher than $10$. As this research focuses on TFC2, the choice was made not to include experiments for values of $K > 10$. To ensure equal lengths between the baseline and the perturbed document the filler token 'a' was used, as it has neutral word semantics and does not change the grammatical structure of the document.

\section{Results}

\subsection{Replication Results}

\begin{figure}[ht]
 \centering
 \begin{minipage}{0.48\textwidth}
 \centering
 \includegraphics[width=\linewidth]{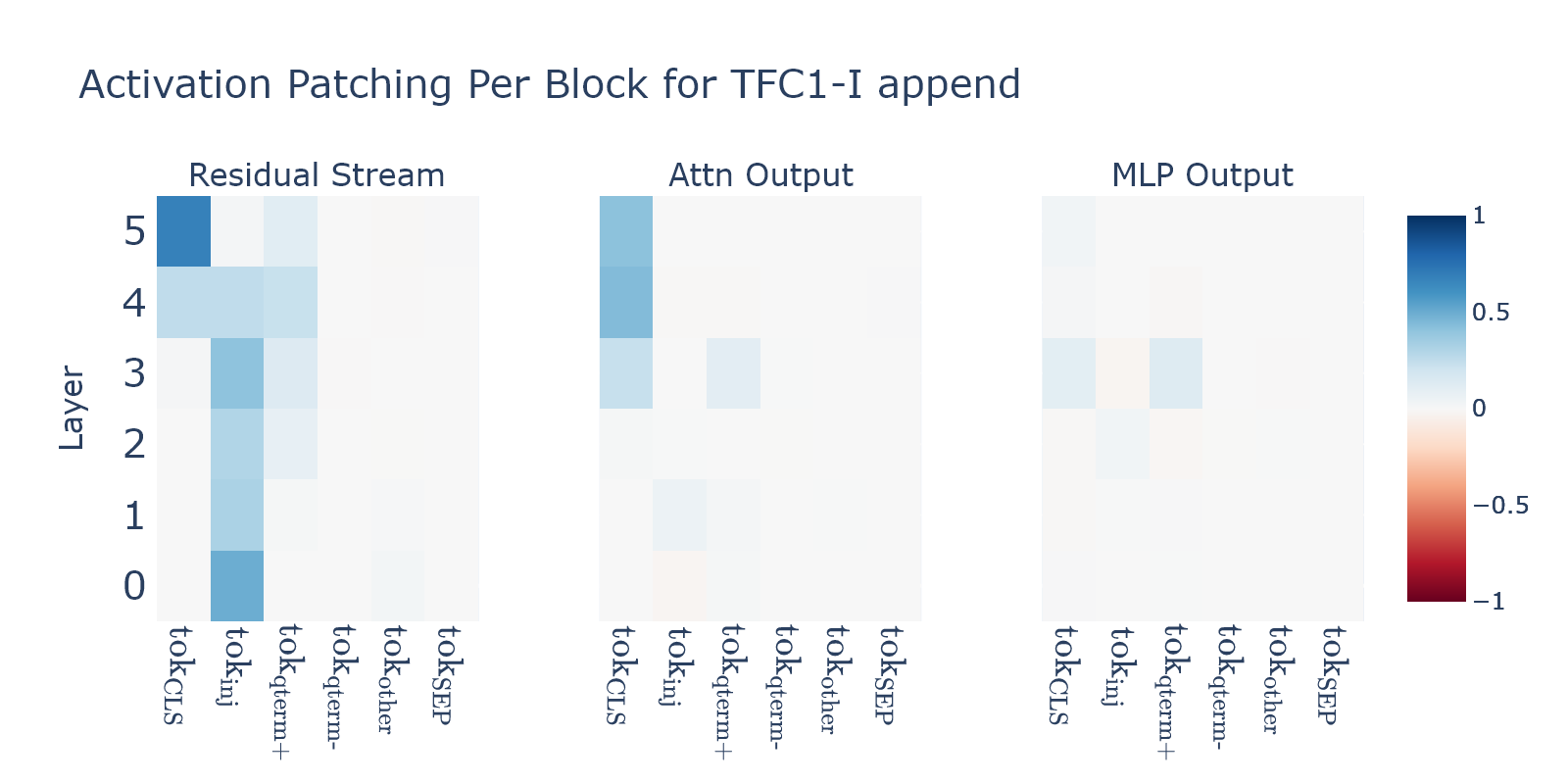}
 \end{minipage}
 \begin{minipage}{0.48\textwidth}
 \centering
 \includegraphics[width=\linewidth]{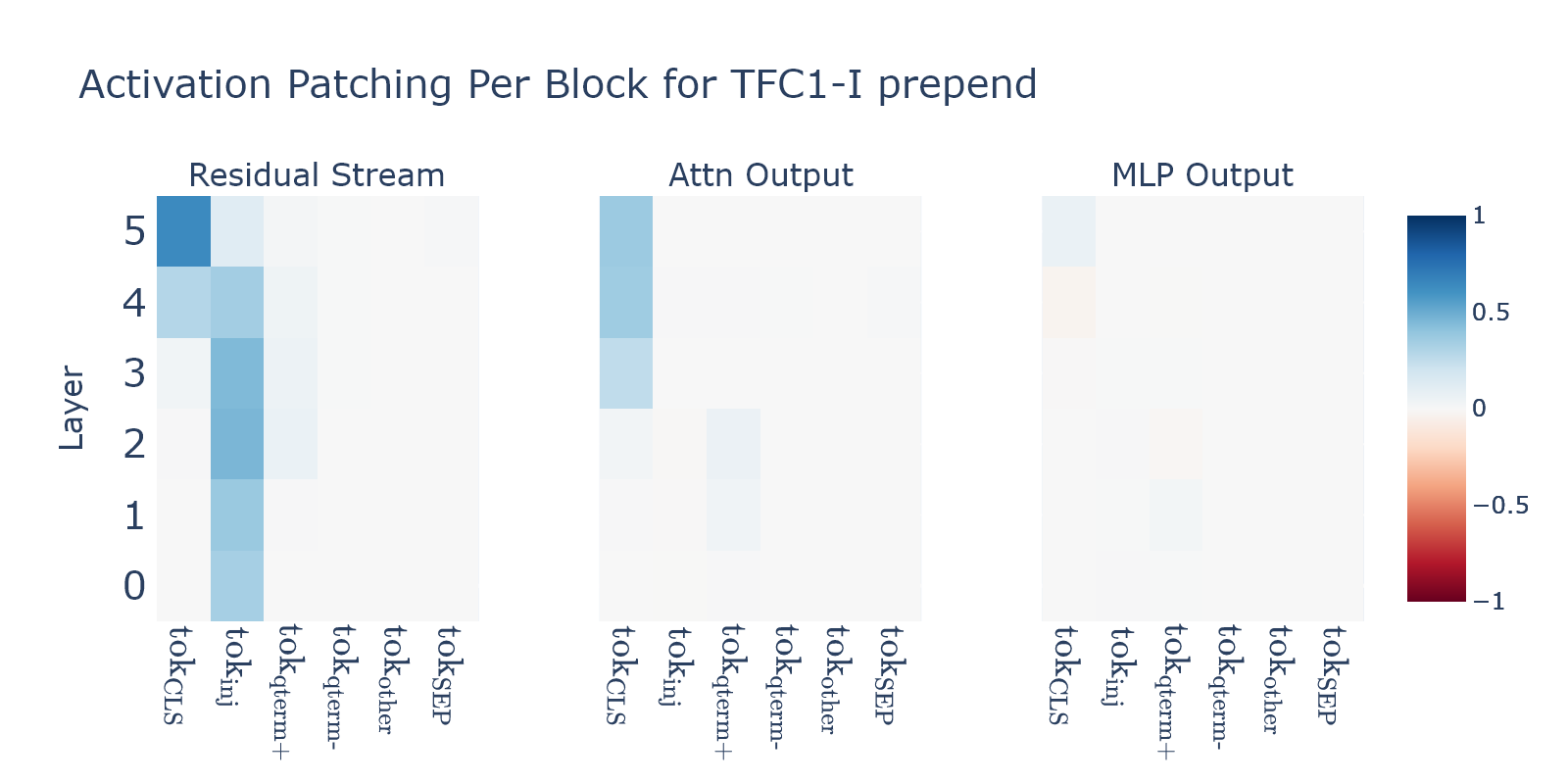}
 \end{minipage}
 \caption{Results of patching the residual stream, attention outputs, and MLP outputs for TFC1-I experiments with append and prepend perturbations. The injected token has the biggest effect on both appended and prepended query terms, while the information is aggregated in the CLS token in latest layers. \protect}
 \Description{Grid of heat maps for patching experiments in the residual stream, attention outputs, and MLP outputs showing which token categories impact the final relevance score. The x-axis contains the token categories. The y-axis denotes the model's layers. A blue color indicates that patching that token class in that layer recovers performance, while a red color denotes the opposite.}
 \label{figure:block}
\end{figure}

As mentioned in Section \ref{section:datasets} and shown in Table \ref{tab:categories}, document tokens are categorized to identify which token types have the greatest impact on model performance across different documents. Figure \ref{figure:block} shows the reproduced results of the TFC1-I append and prepend experiments for patching into the residual stream, the attention outputs, and the MLP outputs at each layer for each token class. A blue color indicates that a token class (x-axis) increases performance when patched in that specific model layer (y-axis). Red indicates the opposite. No color indicates that the patch had no effect.

The first insight is that the term-frequency information is aggregated into the CLS token ($tok_{CLS}$) in layers 4 and 5, aligning with expectations as the ranking score is computed from a pooled representation of the CLS token. In the TFC1-I append experiment, both injected tokens ($tok_{inj}$) and instances of the query term already present in the original document ($tok_{qterm+}$) have an impact on the ranking performance, although the injected tokens play a more significant role. In contrast, when running the TFC1-I prepend experiment, the ranking performance is largely recovered by the injected tokens ($tok_{inj}$) alone. This difference suggests that the model attributes greater importance to the first occurrence of duplicate terms. These results are not in line with the original paper and are discussed in Section \ref{section:discussionbaseline}.

\begin{figure*}[ht]
 \centering
 \includegraphics[width=\linewidth]{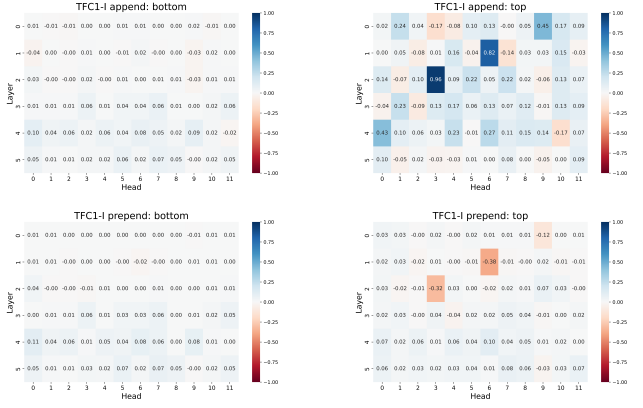}
 \caption{Activation patching on individual attention heads for TFC1-I append and prepend experiments show that term-frequency information is related to attention heads 0.9, 1.6 and 2.3. These attention heads are active when patching the top 10\% ranked documents (right), as opposed to patching the bottom 10\% documents (left). A blue color denotes that patching a specific attention head recovers the perturbed performance, no color denotes that the patch recovers baseline performance, and red indicates that the patch recovers less than the baseline performance. \protect}
 \Description{Four heat maps showing impacts of activation patching per individual attention head. Here a blue color denotes that patching this specific head recovers performance, while a red color denotes the opposite.}
 \label{figure:head_all_TFC1}
\end{figure*}

Following the original paper, activation patching was performed on attention heads to test whether term-frequency information can be localized to specific heads. Figure \ref{figure:head_all_TFC1} presents the results, comparing the top and bottom 10\% ranked documents per query for the TFC1-I experiments. For top-ranked documents in the TFC1-I append experiment, attention heads 0.9 (Layer 0, Head 9), 1.6 and 2.3 are primarily responsible for recovering ranking performance, while their impact on the least relevant documents is minimal. This pattern suggests that these heads reinforce relevance, but do not necessarily indicate it themselves. In the TFC1-I prepend experiment, the same heads (0.9, 1.6, 2.3) influence the relevance score, as shown in Figure \ref{figure:head_all_TFC1}. However, this impact is negative, indicating a fundamental difference in how these heads process term frequency in different configurations. Our analysis finds three attention heads that are relevant to the TFC1 axiom, refining the findings of \citet{chen2024axiomatic}, who identified four attention heads. 

\begin{figure}[ht]
 \centering
 \includegraphics[width=\linewidth]{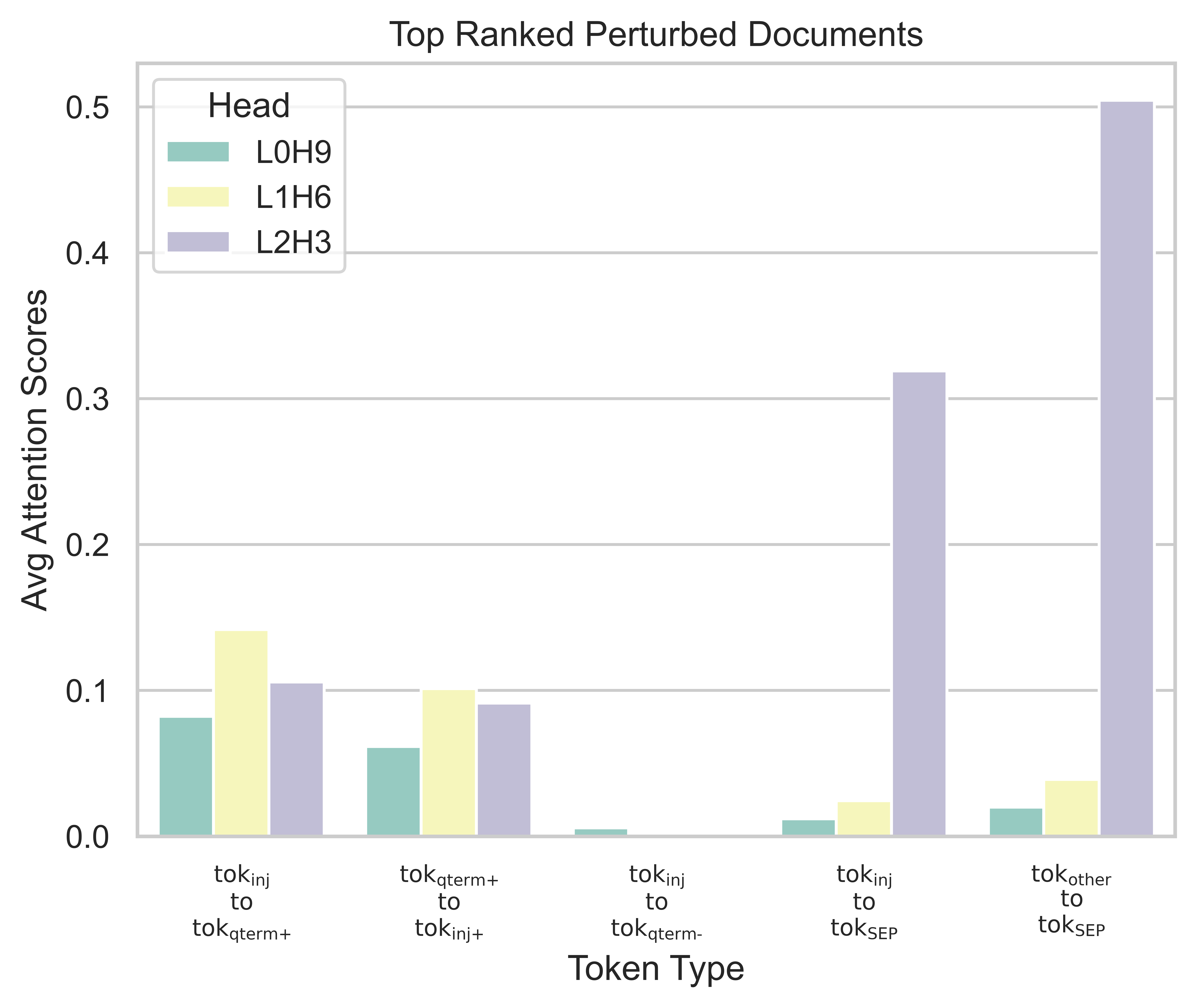}
 \caption{The average attention scores for duplicate token heads show how the most relevant attention heads communicate to construct the relevance signal. More precisely, we see that information gets passed to the separator tokens, especially by node L2H3. \protect}
 \Description{Bar chart demonstrating the contribution of three individual attention heads, by showing the shift between token categories.}
 \label{figure:head_attn}
\end{figure}

Figure \ref{figure:head_attn} illustrates other attention patterns in the TFC1-I append experiments, showing how injected tokens attend to other token categories. We notice that heads 0.9 and 1.6 primarily attend to other occurrences of the selected query term ($tok_{qterm+}$) in the document. In later layers, head 2.3 shifts attention towards the CLS token ($tok_{CLS}$). A similar pattern is observed in the prepend experiment, indicating that term-frequency information initially concentrates on duplicate token occurrences in earlier layers, after which it becomes more diffused across document representations in later layers. This indicates that these heads communicate through the residual stream to compose the relevant signal. This is in line with findings by \citet{chen2024axiomatic}.

\subsection{Extended Results}
The TFC2 axiom states that adding one more query term should increase the ranking score, but the relative difference should become smaller as the frequency of the term in the document increases. However, as the value of $K$ increases, a growing number of document pairs exhibit a counterintuitive trend: the perturbed document (with $K + 1$ added query terms) receives a lower relevance score than the baseline document (with $K$ added query terms). The percentage of document pairs that do not adhere to the TFC1 axiom follows a logarithmic trend, from circa 8\% in TFC2-K1, comparable to the TFC1-I append dataset, to 39\% in TFC2-K10, around which the trend stabilizes. Since TFC2 assumes that TFC1 holds, the results in this section include only query-document pairs that satisfy the TFC1 axiom. The implications of these findings are discussed in Section \ref{section:TFC2implications}.

\begin{figure}[ht]
 \centering
 \includegraphics[width=\linewidth]{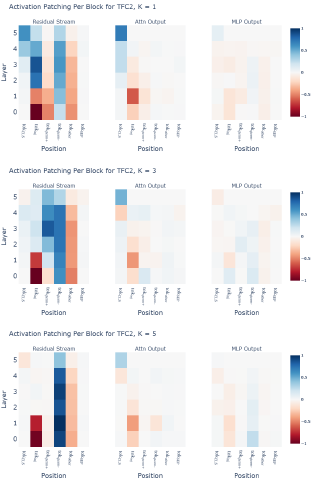}
 \caption{Results of patching the residual stream, attention outputs and MLP outputs for TFC2-K1, TFC2-K3 and TFC2-K5 experiments. We observe that $t_{qterm-}$ has a significant impact for TFC2, while the injected token shows diminishing returns as the frequency of the injected term increases. As with TFC1, information seems to get aggregated in the latest layers under the CLS token. However, the impact is noticeable for a small K. \protect}
 \Description{Grid of heat maps for patching experiments in the residual stream, attention outputs and MLP outputs showing which token categories have an impact on the final relevance score. The x-axis contains the tokens categories. The y-axis denotes the model's layers. A blue color indicates that patching that token class in that layer recovers performance, while a red color denotes the opposite.}
 \label{figure:blockTFC2}
\end{figure}

The results of patching into the residual stream, attention outputs, and MLP outputs for TFC2-K1, TFC2-K3 and TFC2-K5 are shown in Figure \ref{figure:blockTFC2}. Again, a blue color indicates that a token class (x-axis) increases performance when patched in a specific model layer (y-axis). Red indicates the opposite. No color indicates that the patch had no effect.The results for TFC2-K1 are relatively similar to those of TFC1-I (Figure \ref{figure:block}), where the term-frequency information shifts towards the CLS token ($tok_{CLS}$) in later layers, and the injected tokens ($tok_{inj}$) play a significant role in recovering performance. Pre-existing occurrences of selected query terms ($tok_{qterm+}$) previously contributed positively to relevance scores, but now show a slight negative influence. Additionally, query terms not matching the selected query term ($tok_{qterm-}$) now show a minor contribution to model performance. 

This trend continues for higher values of $K$. The model's interest in the injected tokens decreases rapidly, showing a negative impact only when patched in the first layers of the residual stream and attention outputs. The model seems to shift its focus to non-injected instances of the selected query term ($tok_{qterm+}$) and query terms that do not match the selected query term ($tok_{qterm-}$), as seen for TFC2-K3 in Figure \ref{figure:blockTFC2}. For TFC2-K5 this behavior becomes even more apparent, showing that only $tok_{qterm-}$ are important for predicting the relevance score. All experiments with $K > 5$ show the same results, with only $tok_{qterm-}$ being relevant and injected tokens and non-query tokens having a negative influence on performance. These results are in line with the TFC2 axiom: when the frequency of a query term increases, its relevance starts to decrease and the model shows preference for documents with a higher variety of query terms.

\begin{figure*}[!htbp]
 \centering
 \includegraphics[width=0.85\linewidth]{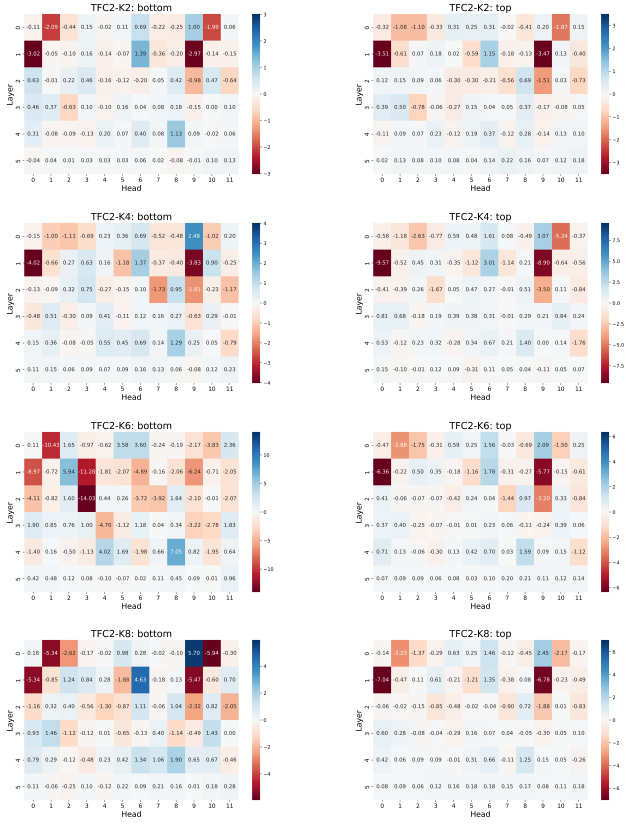}
 \centering
 \caption{Activation patching on individual attention heads for TFC2-K2, TFC2-K4, TFC2-K6, and TFC2-K8, comparing the top and bottom 10\% ranked documents show that attention heads 1.0 and 1.9, as well as 1.6 and 0.9, are related to the TFC2 axiom.A blue color denotes that patching a specific attention head recovers the perturbed performance, no color denotes that the patch recovers baseline performance, and red indicates that the patch recovers less than the baseline performance. \protect}
 \Description{Multiple heat maps showing the effect of individual attention heads in top and bottom ranked documents. A blue color denotes that patching this specific attention head recovers original performance, while a red color denotes the opposite.}
 \label{figure:head_all_TFC2}
\end{figure*}

Patching attention heads for TFC2-K1 through TFC2-K10, shown in Figure \ref{figure:head_all_TFC2}, reveals that attention heads 1.0 and 1.9 strongly influence ranking performance negatively, both in the top and bottom 10\% of ranked documents. Heads 1.6 and 0.9 show a mostly positive influence on ranking performance. Unlike the TFC1-I experiments, attention heads 1.0 and 1.9 are relevant in both the relevant top and bottom documents.

\begin{figure}[ht]
 \centering
 \includegraphics[width=\linewidth]{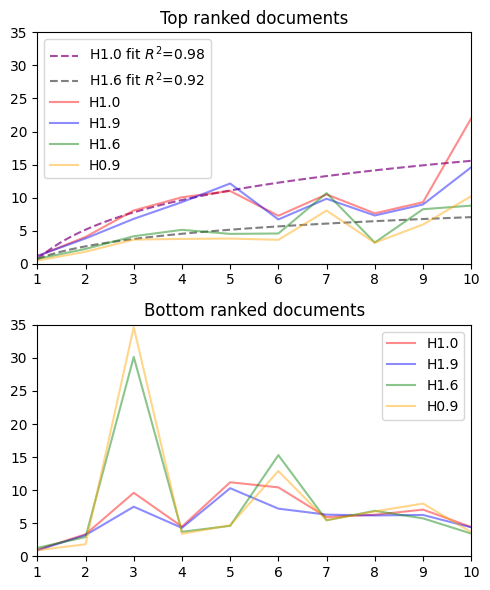}
 \caption{Average absolute values of attention heads 1.0, 1.9, 1.6 and 0.9 for the top and bottom ranked documents. The average attention scores for head 1.0 and 1.6 for top ranked documents are plotted for different values of $K$ and show a sublinear trend. \protect}
 \Description{Three line plots describing the average impact of two attention heads.}
 \label{figure:specificheads}
\end{figure}

Figure \ref{figure:specificheads} shows the average values of patching heads 1.0, 1.9, 1.6 and 0.9. The reported values are absolute, as we are mainly interested in the magnitude of the impact rather than its direction. The most noticeable observation is that heads 1.0 and 1.9 show very similar behavior, as well as heads 1.6 and 0.9. This indicates that the pairs of heads might either store redundant information, reinforce a copying pattern, or encode a single function across both heads. Similar behavior has been seen in research by \citet{wang2022interpretability}.

For the top 10\% of ranked documents documents, the plot shows values that increase with $K$. For the bottom 10\% of ranked documents, we see a fluctuating trend, both for heads 1.0 and 1.9 as for heads 1.6 and 1.9. What stands out for the top ranked documents is that for $K > 5$, the values start to fluctuate. This can be explained by behavior that also became apparent in Figure \ref{figure:blockTFC2}: after $K = 5$, the model is not interested in the injected tokens anymore, and solely focuses on query terms other than the selected one. 

To demonstrate that the heads active for TFC2 behave in line with the axiom, we will show that their values follow a sublinear relation for $K$ values until 5. Averaging the impact of attention heads 1.0 and 1.9 over the top ranking documents reveals that the observed trend follows a logarithmic function $a \log(x) + b$ with an $R^2$ value of $0.98$, where $a=6.47$ and $b=0.67$. Averaging the impact of attention heads 1.6 and 0.9 over the top ranking documents reveals that the observed trend follows a logarithmic function $a \log(x) + b$ with an $R^2$ value of $0.92$, where $a=2.75$ and $b=0.73$. We can show that these logarithmic functions $g(x)$ are guaranteed to be sublinear by the following derivation: 

To show that \( g(x) = a \log(x) + b \) is sublinear, we must show it grows asymptotically slower compared to \( f(x) = c x + d \). This is equivalent to showing the following limit approaches zero:

\begin{equation}
  \lim_{x \to \infty} \frac{g(x)}{f(x)} = \lim_{x \to \infty} \frac{a \log(x) + b}{c x + d}
\end{equation}

We apply l'Hôpital since this is a $\frac{\infty}{\infty}$ limit: 

\begin{equation}
  \lim_{x \to \infty} \frac{g(x)}{f(x)} = \lim_{x \to \infty} \frac{a \frac{1}{x}}{c} = \frac{a}{c}\lim_{x \to \infty} \frac{1}{x} = 0
\end{equation}

This result is frequently expressed in the mathematical literature using little-o notation, which formally states that \(g(x) = o(f(x)) \text{ as } x \to \infty.\)

\section{Discussion}
In this work, we investigate the claims of \citet{chen2024axiomatic} on encoding term frequency in neural ranking models, and introduce new experiments for the TFC2 axiom.

\subsection{Implications of the Diagnostic Datasets}
\label{section:implicationsdataset}
\citet{chen2024axiomatic} define practices to create a diagnostic dataset for activation patching experiments in an IR setting. As described in Section \ref{section:datasets}, the 100 queries with the highest change in the average retrieval score are selected to create the diagnostic dataset. Therefore, the queries used for the activation patching experiment are task dependent. Nevertheless, the initial experiments for both append and prepend variants have been run on the same query dataset. As a result, this study also uses the queries from the original experiments. The selected query terms have also been adopted from the original experimental setup.

The TFC1-R perturbation as defined by \citet{chen2024axiomatic} involves replacing all instances of a selected query term with a filler term. However, we found that only $1837$ replacements were made in $1029$ documents (\~10\% of all documents), which results in many instances where the baseline and perturbed documents were identical. This effectively reduces the number of instances that isolate the feature we want to observe. Therefore, the results of this experiment are noisy and not reliable. Furthermore, since the ranking metric (Section \ref{section:experimentalTFC1}) is normalized by the difference between the perturbed and baseline scores, these minimal differences lead to unstable calculations. Therefore, we have chosen to leave the TFC1-R experiments out of this study.

\subsection{Implications of the Baseline Setup}
\label{section:discussionbaseline}
Although we have been able to validate the main claims made by \citet{chen2024axiomatic}, there are some discrepancies in the reproduced results due to a difference in the baseline setup for the TFC1-I append experiments. While described correctly in the original paper, the code base contained baseline documents with a filler token in a different location than specified.

Specifically, when patching with the corrected baseline (Figure \ref{figure:block}), we observe a shift in performance recovery from instances of $tok_{qterm+}$ to $tok_{inj}$, although $tok_{qterm+}$ do retain some influence. When looking at individual attention heads (Figure \ref{figure:head_all_TFC1}), the results broadly align with the original paper's findings, with a slight change in which attention heads have the biggest impact. While the original study identified heads 0.9, 1.6, 2.3 and 3.8 as containing term-frequency information, our analysis more specifically shows that heads 0.9, 1.6 and 2.3 are responsible for this behavior. This improvement in baseline setup has enabled us to more precisely identify the internal mechanisms responsible for term-frequency tracking, narrowing it down to three specific attention heads (0.9, 1.6, and 2.3) rather than the four heads identified in the original study. 

\subsection{Implications of TFC2 Experiments}
\label{section:TFC2implications}
The TFC2 axiom is built upon TFC1, stating that while the relevance score increases when more occurrences of a query term are added (TFC1), this increase follows a sublinear trend, as was shown in Figure \ref{figure:specificheads}. Analysis of the TFC2 diagnostic datasets across different values for $K$ revealed an unexpected pattern: an increasing portion of document pairs did not adhere to the TFC1 axiom (perturbed score < baseline score). This rate grew logarithmically with $K$, from around 8\% of document pairs for TFC1-K1, which is comparable with the TFC1-I append dataset, to a plateau around 39\% for TFC1-K50. All query-document pairs not adhering to TFC1 have been filtered from the results, which greatly reduced noise and fluctuations.

Our TFC2 experiments rely on diagnostic datasets that are constructed by appending a query term multiple times, which introduces several issues. First, this method generates documents with unnatural term repetitions that likely fall outside the model's training distribution. Second, as $K$ increases, repeatedly appending the query term alters the token distribution, potentially flattening the probability distribution over all tokens. Third, the difference between the ranking score of the baseline and perturbed documents becomes smaller as values of $K$ increase, leading to numerical instability. 

The experiments with patching the residual stream, attention outputs, and MLP outputs (Figure \ref{figure:blockTFC2}) show that the model reacts to these unnatural abundance of a query term by discarding that information and focusing more on other query terms, which is in line with TFC2. However, a more sophisticated approach than appending a query term multiple times could give a more practical into the inner working of a retrieval model. We will address in the following section.

\subsection{Future Research}
This research demonstrates that neural ranking models not only are complex to interpret but also deviate from fundamental axiomatic properties of information retrieval. Opposed to the research by \citet{chen2024axiomatic}, experiments on diagnostic datasets show that neural ranker models do not consistently adhere to the TFC1 axiom. When there is adherence to TFC1, we are able to show that the model shows behavior in line with TFC2.

Our findings highlight the importance of carefully curated diagnostic datasets for activation patching experiments. \citet{chen2024axiomatic} mentioned the importance of perturbation location, demonstrated by the difference in the TFC1-I append and prepend experiments. The current approach of selecting a single query term per query and applying it to the top 100 relevant documents might be oversimplified, as it fails to account for document-specific term interactions and context. 

Several directions for future research emerge from these findings. First, diagnostic dataset creation could be improved by incorporating document-specific query term selection, instead of perturbing all retrieved documents for one query with the same query term. Second, the influence of term repetition on neural ranker models could use investigation, as the current diagnostic datasets contain unnatural text. Finally, these observed deviations call for the need for more in-depth research into the presence of axioms in the ranking model's behavior.

\section{Conclusion}
This paper reproduced and extended research by \citet{chen2024axiomatic}, which introduced an activation patching method for IR. \citet{chen2024axiomatic} found that neural ranking models adhere to the TFC1 axiom and were able to trace this behavior back to a specific group of attention heads. Although this framework can be used to track behavior within neural ranker models, this reproducibility study shows that the results strongly depend on the design of diagnostic datasets. By extending the framework to include the TFC2 axiom - a generalization of the TFC1 axiom - we aimed to gain more insight into the inner decision process of neural ranking models. Contrary to initial expectations, our findings show that ranking scores do not consistently adhere to the TFC1 axiom when more query terms are injected. Despite this, we have been able to identify attention heads that are highly relevant to and behave in line with the TFC2 axiom. This study is a next step in interpretability efforts within IR and demonstrates that neural ranking models might be less rooted in interpretable axioms than expected, highlighting the need for continued research into model transparency and behavior.

\bibliographystyle{ACM-Reference-Format}
\balance
\bibliography{bibliography}

\appendix
\end{document}